\documentclass[reprint,superscriptaddress,amsmath,amssym,aps,prb]{revtex4-2}

\usepackage{bm}
\usepackage{float}
\usepackage{graphicx}
\usepackage{mathtools}
\usepackage{braket}
\usepackage[usenames,dvipsnames]{color}
\usepackage[normalem]{ulem}
\usepackage[svgnames]{xcolor}
\usepackage{bm}
\usepackage{multirow}
\usepackage{titlesec}
\usepackage[utf8]{inputenc}
\usepackage{array}
\usepackage[colorlinks,linkcolor=blue,citecolor=blue,urlcolor=blue]{hyperref}

\begin{document}
\title{Comment on `High-resolution Measurements of Thermal Conductivity Matrix and Search for Thermal Hall Effect in La$_2$CuO$_4$' }

\author{Shan Jiang}
\affiliation{Laboratoire de Physique et d'Etude de Mat\'{e}riaux (CNRS)\\ ESPCI Paris, PSL Research University, 75005 Paris, France }
\affiliation{Wuhan National High Magnetic Field Center and School of Physics, Huazhong University of Science and Technology,  Wuhan  430074, China}

\author{Qiaochao Xiang}
\affiliation{Wuhan National High Magnetic Field Center and School of Physics, Huazhong University of Science and Technology,  Wuhan  430074, China}
\author{Beno\^it Fauqu\'e}
\affiliation{Laboratoire de Physique et d'Etude de Mat\'{e}riaux (CNRS)\\ ESPCI Paris, PSL Research University, 75005 Paris, France }

\author{Xiaokang Li}
\affiliation{Wuhan National High Magnetic Field Center and School of Physics, Huazhong University of Science and Technology,  Wuhan  430074, China}

\author{Zengwei Zhu}
\affiliation{Wuhan National High Magnetic Field Center and School of Physics, Huazhong University of Science and Technology,  Wuhan  430074, China}

\author{Kamran Behnia} 
\affiliation{Laboratoire de Physique et d'Etude de Mat\'{e}riaux (CNRS)\\ ESPCI Paris, PSL Research University, 75005 Paris, France }
\begin{abstract}
Recently, Jiayi Hu and co-workers reported that they did not resolve any thermal Hall signal in  La$_2$CuO$_4$ by `high resolution' measurements, setting an upper bound of $|\kappa_{xy}| <2\times 10^{-3}~$Wm$^{-1}$K$^{-1}$ at 20 K. Two points have apparently escaped their attention. First, thermal Hall signals with an amplitude well below this resolution bound have been detected in disordered perovskites. Second, the longitudinal thermal conductivity of their sample is significantly lower than the La$_2$CuO$_4$ sample displaying a thermal Hall signal. We find that a moderate reduction of $\kappa_{xx}$ in SrTiO$_3$ is concomitant with a drastic attenuation of $\kappa_{xy}$. A trend emerges across several families of insulators: the amplitude of $\kappa_{xy}$ anti-correlates with disorder. 
\end{abstract}
\maketitle

 \begin{figure}[ht!]
\begin{center}
\centering
\includegraphics[width=8.5cm]{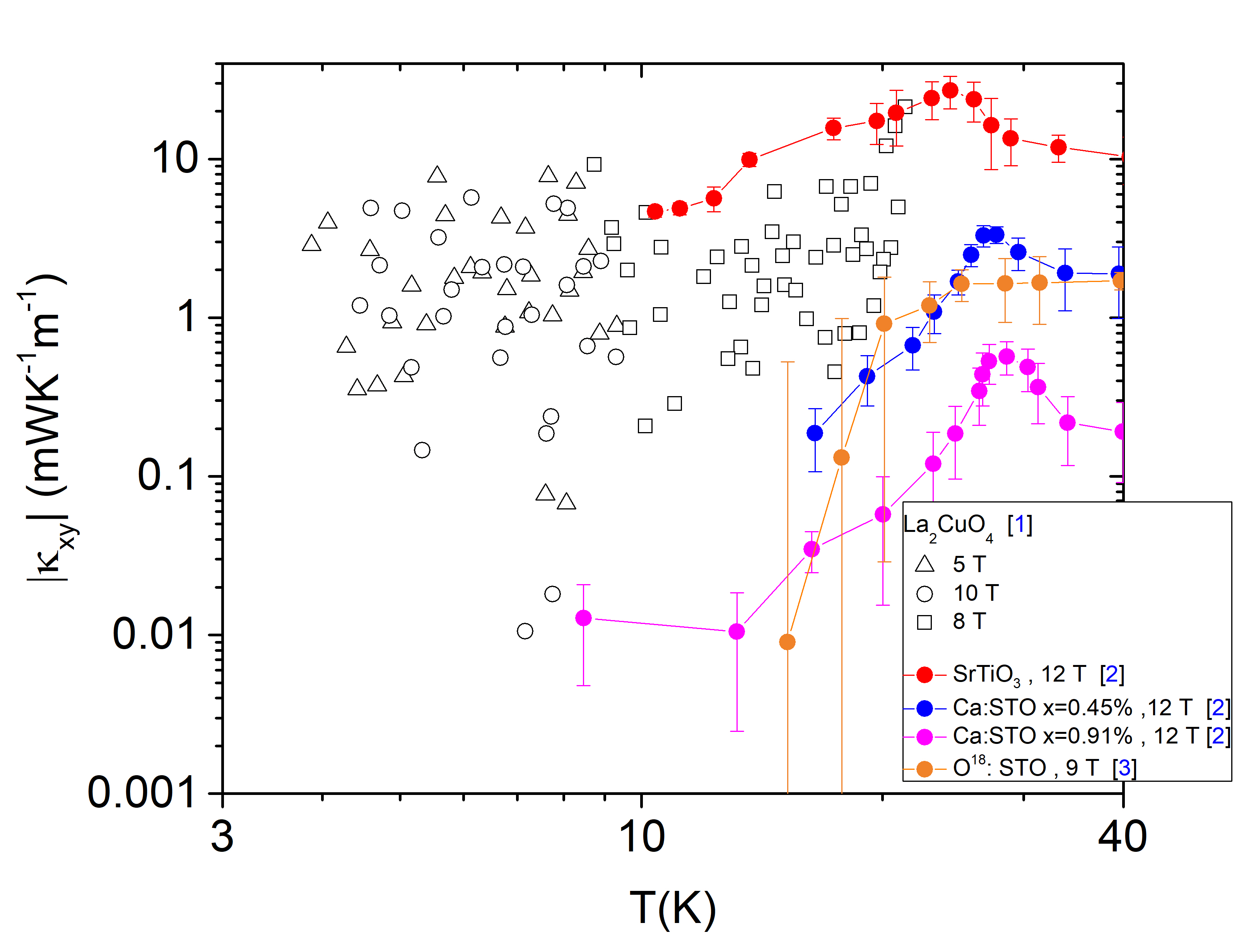} 
\caption{Empty symbols represent  $|\kappa_{xy}|$ data gathered between 2 K and 20 K at three magnetic fields reported in ref. \cite{hu2025highresolutionmeasurementsthermalconductivity}. Red circles show $-\kappa_{xy}$ in nominally pure SrTiO$_3$ reported in ref. \cite{Jiang2022}. Blue and pink circles show $-\kappa_{xy}$ in Sr$_{0.9955}$Ca$_{0.0045}$TiO$_3$ and in Sr$_{0.9909}$Ca$_{0.0091}$TiO$_3$\cite{Jiang2022}. Orange circles represent $-\kappa_{xy}$ in a SrTiO$_3$ crystal in which most $^{16}$O were replaced by $^{18}$O \cite{Sim2021}. The noise level of the data in \cite{hu2025highresolutionmeasurementsthermalconductivity} is too large to detect a $\kappa_{xy}$ as small as what was found in these studies.}
\label{fig1}
\end{center}
\end{figure}

In a recent post, Hu \textit{et al.} \cite{hu2025highresolutionmeasurementsthermalconductivity} report on an investigation of longitudinal and Hall conductivity in a La$_2$CuO$_4$ crystal. In sharp contrast with previous reports \cite{Grissonnanche2019,Grissonnanche2020,Boulanger2020}, they do not find any detectable thermal Hall signal.

Thermal Hall effect, which refers to the generation of a transverse thermal gradient by a longitudinal heat current, was first reported to be caused by phonons in an insulator two decades ago \cite{Strohm2005}. In recent years, it has been detected in numerous insulators belonging to a diverse variety of insulators including multiferroics \cite{Ideue2017},  perovskites \cite{Li2020,Sim2021,Jiang2022}, spin ice \cite{Uehara2022}, topological insulators \cite{Sharma2024}, honeycomb antiferromagnets \cite{Kasahara2018,Hentrich2019,Lefran2022,Bruin2022,Meng2024} and even common elemental solids \cite{Li2023,Jin2024}.  The results reported by Hu \textit{et al.} \cite{hu2025highresolutionmeasurementsthermalconductivity}appears to raise two question: How solid is the experimental consensus on the ubiquity of this effect? How reproducible are the thermal Hall measurements signal? 

Therefore, their paper deserves careful scrutiny. 

The first point which attracts attention is the presence of the adjective `high-resolution' in the title of the paper. The method employed is a standard one-heater-three-thermometer set-up widely used by other researchers. As for resolution, it is surprisingly low.  Figure \ref{fig1} compares the noise level reported in Figure 2b of ref. \cite{hu2025highresolutionmeasurementsthermalconductivity} with the data reported by two other groups on strontium titanate \cite{Jiang2022,Sim2021}. One can see that the noise level of the set-up used in ref. \cite{hu2025highresolutionmeasurementsthermalconductivity} is unusually large. Such a noise level would have allowed the detection of a thermal Hall signal in nominally pure strontium titanate, where $\kappa_{xy} \approx 20 ~$mWK$^{-1}$m$^{-1}$ \cite{Li2020,Sim2021,Jin2024}, but not the one in strontium titanate subject to atomic substitution \cite{Sim2021,Jiang2022} ($\kappa_{xy} \approx 1 ~$mWK$^{-1}$m$^{-1}$).  
 
 \begin{figure}[ht!]
\begin{center}
\centering
\includegraphics[width=8.5cm]{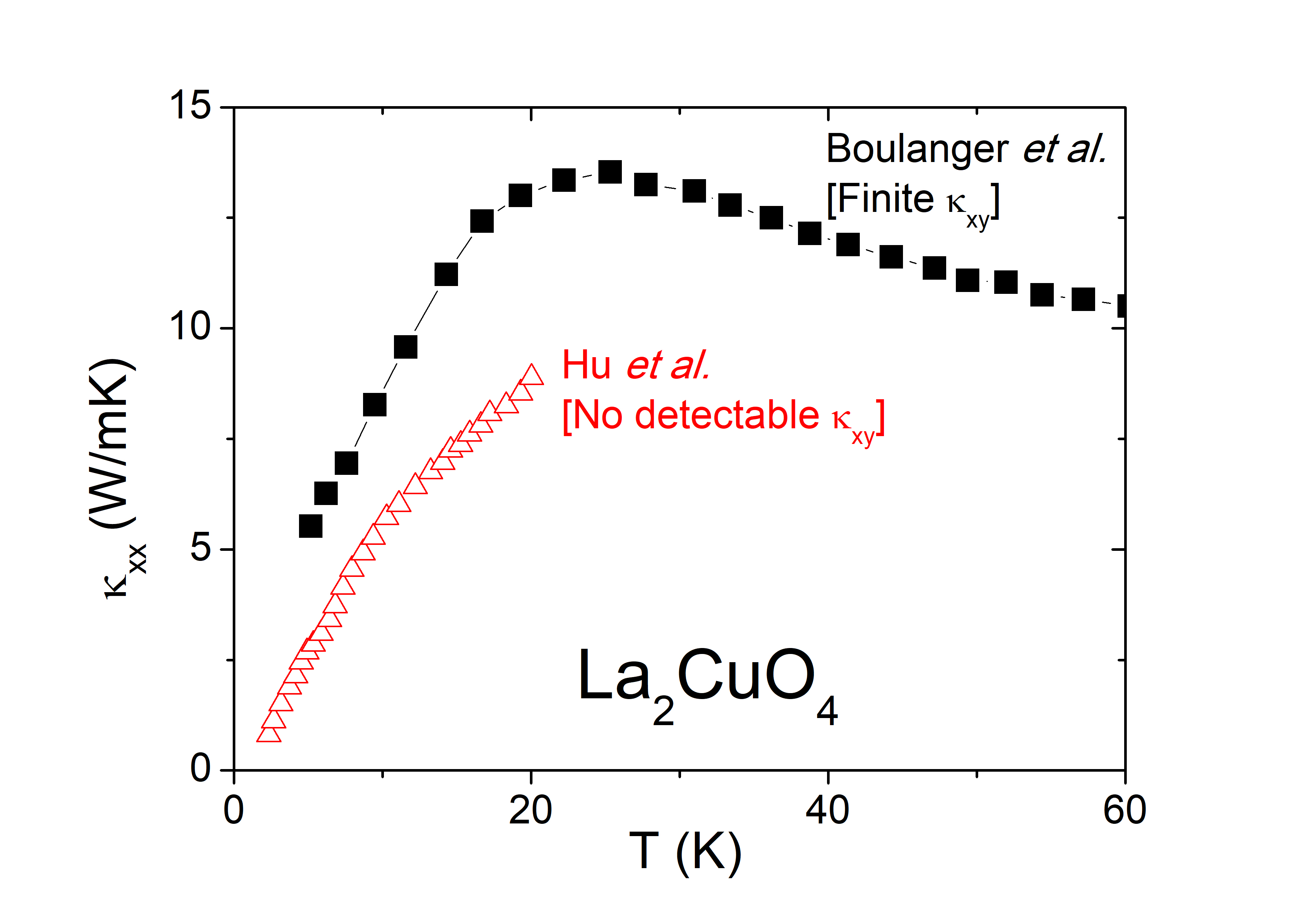} 
\caption{A comparison of longitudinal thermal conductivity in La$_2$CuO$_4$ reported in \cite{hu2025highresolutionmeasurementsthermalconductivity} and in \cite{Boulanger2020}. }
\label{fig2}
\end{center}
\end{figure}

Several authors had previously employed the adjective `giant' to designate what they have measured as a thermal Hall  signal. This semantic choice neglects the fact that the latter is more than a hundred times smaller than the longitudinal response. Few, if any, had advertised the resolution of the measurements before Hu \textit{et al} and their remarkable noise level.

 \begin{figure}[ht!]
\begin{center}
\centering\includegraphics[width=8.5cm]{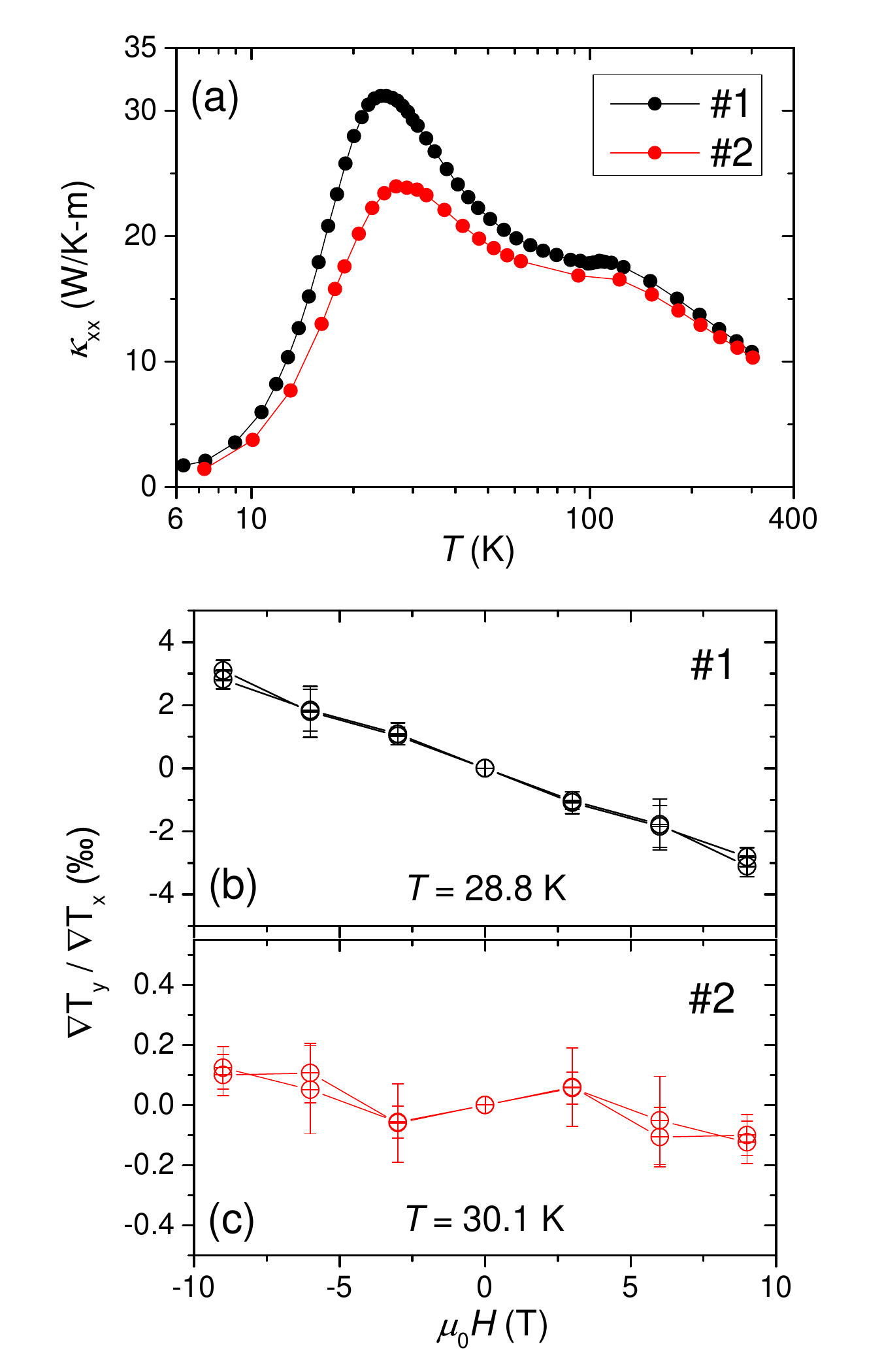} 
\caption{a) Longitudinal thermal conductivity in two SrTiO$_3$ crystals. The peak thermal conductivity in sample $\#1$ is 1.3 times larger than in sample $\#2$. b) The ratio of transverse to longitudinal temperature gradient near the peak temperature is sample $\#1$. There is a clear signal. c) Same for  sample $\#2$. There is no detectable signal. The dirtier sample has a somewhat lower $\kappa_{xx}$ and a drastically reduced $\kappa_{xy}$. }
\label{fig3}
\end{center}
\end{figure}

The second point deserving attention is in the abstract \cite{hu2025highresolutionmeasurementsthermalconductivity}, where they write: ``The longitudinal thermal conductivity $\kappa_{xx}/T$ agrees well with previous studies, in both magnitude and T dependence." Figure \ref{fig2} compares the zero-temperature longitudinal thermal conductivity of La$_2$CuO$_4$ reported in ref. \cite{Boulanger2020} with what was measured by Hu {et al.} \cite{hu2025highresolutionmeasurementsthermalconductivity}. The discrepancy between the two sets of data has apparently escaped their attention.  At T = 20 K, the thermal conductivity of the sample which hosted an easily measurable $\kappa_{xy}$ is 1.4 times larger than the sample which did not.

The amplitude of the peak thermal conductivity is commonly used as a criterion for purity \cite{kawabata2025}. Isotopic purification is known to generate a very large enhancement of the peak thermal conductivity in diamond \cite{Wei1993} and in silicon \cite{Inyushkin} . It is therefore legitimate to speculate that the striking difference in the amplitude of the thermal Hall response in the two contradicting reports is simply because in cleaner La$_2$CuO$_4$ crystals $\kappa_{xy}$ is drastically larger than in  dirtier ones. 

Such an anti-correlation between the amplitude of $\kappa_{xy}$ and disorder is compatible with what is known in other insulators. In the case of NiPS$_3$, the amplitude of $\kappa_{xy}$ is strongly affected by the presence of structural domains, which reduce the magnetization anisotropy \cite{Meng2024}. A correlation between amplitudes of $\kappa_{xy}$ and $\kappa_{xx}$ is also visible in the data regarding $\alpha$-RuCl$_3$ (another antiferromagnet with a honeycomb lattice). The data was compiled to demonstrate a correlation between sample quality and the presence of what is dubbed, thanks to a generous stretch of imagination, a thermal Hall `plateau' \cite{Kasahara2022}. 

The most illuminating case, however, is strontium titanate. As seen in Figure~\ref{fig1}, substituting a tiny fraction ($<0.005$) of Sr atoms by Ca atoms in SrTiO$_3$ attenuates $\kappa_{xy}$ by more than one order of magnitude \cite{Jin2024}. A similar attenuation is achieved by replacing most $^{16}$O atoms with $^{18}$O. This indicates a drastic effect of disorder on $\kappa_{xy}$. We have also found that even in nominally pure strontium titanate there is a strong anti-correlation between disorder and the amplitude of $\kappa_{xy}$.  Figure \ref{fig3} shows the data in two nominally identical SrTiO$_3$ crystals, commercially obtained from two different companies. One can see that a moderate attenuation of $\kappa_{xx}$ (of the order of the difference in the $\kappa_{xx}$ of the two La$_2$CuO$_4$ crystals) is sufficient to make $\kappa_{xy}$ undetectably small. The anti-correlation between disorder and the amplitude of $\kappa_{xy}$ in SrTiO$_3$ will be the subject of a future detailed study.  However, as one can see in table \ref{Table_1}, available data indicates that $\kappa_{xy}$ falls drastically when the peak of $\kappa_{xx}$ falls below $\approx 25$ WK$^{-1}$m$^{-1}$.

In summary,  Hu \textit{et al.} overlooked that the amplitude of $\kappa_{xy}$ detected in disordered perovskites is below their experimental resolution and that their sample was more disordered than the cuprate sample previously studied.  Recognition of these features would drastically modify the implications of their findings. 

\begin{table*}[ht!]
\centering
\begin{tabular}{|c|c|c|c|c|c|c|}
\hline
City of measurement & $\kappa_{xx}^{peak}$(WK$^{-1}$m$^{-1}$) & $-\kappa_{xy}^{peak}$(mWK$^{-1}$m$^{-1}$)&$B_{max}$(T)& $T_{peak}$ (K) & Reference \\
\hline
\hline
Paris & 36 & 80& 12&21 &\cite{Li2020} \\
Paris & 35  & 27& 12&25 &\cite{Jiang2022} \\
Paris & 32  & 21&12&25 &\cite{Jiang2022} \\
Seoul & 33  & 96 &9&25 &\cite{Sim2021} \\
Seoul & 23  &2 &9&28 &\cite{Sim2021} \\
Shanghai & 39  & 29 &9 &23 &\cite{Jin2024} \\
Wuhan& 31  & $90$&9 &25 & To be published \\
Wuhan& 24  & $2$ &9 &27 & To be published \\

\hline
\end{tabular}
\caption{ The thermal conductivity data on SrTiO$_3$ reported in different places by different groups. While the order of magnitude of $\kappa_{xy}$ is similar for samples with a reasonably large $\kappa_{xx}$, it drastically drops in samples with a pathologically low thermal conductivity.}
\label{Table_1}
\end{table*}

\bibliography{main.bib}

\end{document}